\documentclass[conference]{IEEEtran}
\IEEEoverridecommandlockouts
\usepackage{tikz}

\newcommand\copyrighttext{%
  \normalfont\footnotesize \textcopyright~2026 IEEE. Personal use of this material is permitted.
  Permission from IEEE must be obtained for all other uses, in any current or future
  media, including reprinting/republishing this material for advertising or promotional
  purposes, creating new collective works, for resale or redistribution to servers or
  lists, or reuse of any copyrighted component of this work in other works.
  DOI: \href{https://doi.org/10.1109/COMPSAC69091.2026.00015}{10.1109/COMPSAC69091.2026.00015}}
\newcommand\copyrightnotice{%
\begin{tikzpicture}[remember picture,overlay]
\node[anchor=south,yshift=27.5pt] at (current page.south) {\parbox{\textwidth}{\copyrighttext}};
\end{tikzpicture}%
}

\usepackage{cite}
\usepackage{amsmath,amssymb,amsfonts}
\usepackage{algorithmic}
\usepackage{graphicx}
\usepackage{textcomp}
\usepackage{xcolor}
\usepackage[utf8]{inputenc}
\usepackage[english]{babel}
\usepackage{multirow}
\usepackage{booktabs}
\usepackage{tabularx}
\usepackage{array}
\usepackage{url}
\usepackage{tcolorbox}
\usepackage{listings}
\usepackage{enumitem} 
\usepackage{microtype}
\usepackage{bookmark}
\usepackage{todonotes}
\usepackage[caption=false,font=footnotesize]{subfig}
\usepackage{url}

\usepackage{soul} 

\usepackage[binary-units=true]{siunitx}

\usepackage[caption=false, font=footnotesize]{subfig}

\definecolor{darkblue}{rgb}{0.0,0.0,0.3} 
\definecolor{darkred}{rgb}{0.4,0.0,0.0}
\definecolor{red}{rgb}{0.7,0.0,0.0}
\definecolor{lightgrey}{rgb}{0.8,0.8,0.8} 
\definecolor{grey}{rgb}{0.6,0.6,0.6}
\definecolor{darkgrey}{rgb}{0.4,0.4,0.4}

\usepackage[acronym, nopostdot]{glossaries}
\glsdisablehyper
\global\defaultaddspace=3pt

\definecolor{codegreen}{rgb}{0,0.6,0}
\definecolor{codegray}{rgb}{0.5,0.5,0.5}
\definecolor{codepurple}{rgb}{0.58,0,0.82}
\definecolor{backcolour}{rgb}{0.95,0.95,0.92}

\lstdefinelanguage{json}{
    basicstyle=\ttfamily\footnotesize,
    backgroundcolor=\color{backcolour},
    stringstyle=\color{codepurple},
    identifierstyle=\color{black},
    keywordstyle=\color{blue},
    numbers=left,
    numberstyle=\tiny\color{codegray},
    breaklines=true,
    frame=single,
    showstringspaces=false,
    literate=
     *{0}{{{\color{blue}0}}}{1}
      {1}{{{\color{blue}1}}}{1}
      {2}{{{\color{blue}2}}}{1}
      {3}{{{\color{blue}3}}}{1}
      {4}{{{\color{blue}4}}}{1}
      {5}{{{\color{blue}5}}}{1}
      {6}{{{\color{blue}6}}}{1}
      {7}{{{\color{blue}7}}}{1}
      {8}{{{\color{blue}8}}}{1}
      {9}{{{\color{blue}9}}}{1}
      {:}{{{\color{black}{:}}}}{1}
      {,}{{{\color{black}{,}}}}{1}
      {\{}{{{\color{black}{\{}}}}{1}
      {\}}{{{\color{black}{\}}}}}{1}
      {[}{{{\color{black}{[}}}}{1}
      {]}{{{\color{black}{]}}}}{1},
}
\newacronym{AES}{AES}{accuracy-efficiency score}
\newacronym{AI}{AI}{artificial intelligence}
\newacronym{API}{API}{application programming interface}
\newacronym{AWS}{AWS}{Amazon Web Services}
\newacronym{CI}{CI}{confidence interval}
\newacronym{COP}{CoP}{cost-of-pass}
\newacronym{CPSQ}{CpSQ}{cost per successful query}
\newacronym{CPU}{CPU}{central processing unit}
\newacronym{CV}{CV}{coefficient of variation}
\newacronym{DMAS}{DMAS}{distributed multi-agent system}
\newacronym{EBS}{EBS}{Elastic Block Storage}
\newacronym{GUI}{GUI}{graphical user interface}
\newacronym{IDK}{IDK}{i don't know}
\newacronym{IO}{I/O}{input/output}
\newacronym{IoA}{IoA}{internet of agents}
\newacronym{JSON}{JSON}{JavaScript object notation}
\newacronym{KV}{KV}{key value}
\newacronym{LLM}{LLM}{large language model}
\newacronym{LTS}{LTS}{long-term support}
\newacronym{LTM}{LTM}{long-term memory}
\newacronym{MAS}{MAS}{multi-agent system}
\newacronym{MBIT}{Mbit}{megabit}
\newacronym{MBIT/S}{Mbit/s}{megabit per second}
\newacronym{MBITS}{Mbits}{megabits}
\newacronym{MBITS/S}{Mbits/s}{megabits per second}
\newacronym{MBYTE}{MByte}{megabyte}
\newacronym{MBYTE/S}{MBytes/s}{megabyte per second}
\newacronym{MBYTES}{MBytes}{megabytes}
\newacronym{MBYTES/S}{MBytes/s}{megabyte per second}
\newacronym{OS}{OS}{operating system}
\newacronym{QA}{Q\&A}{questions and answers}
\newacronym{RAG}{RAG}{retrieval-augmented generation}
\newacronym{RAM}{RAM}{random access memory}
\newacronym{RQ}{RQ}{research question}
\newacronym{SD}{SD}{standard deviation}
\newacronym{SLM}{SLM}{small language model}
\newacronym{TCO}{TCO}{total cost of ownership}
\newacronym{TCP}{TCP}{transmission control protocol}
\newacronym{USD}{USD}{United States Dollar}
\newacronym{VCPU}{vCPU}{virtual central processing unit}
\newacronym{WSL}{WSL}{Windows Subsystem for Linux}

\begin{document}

\title{Cost and Accuracy of Long-Term Memory in Distributed Multi-Agent Systems Based on Large Language Models}

\author{\IEEEauthorblockN{Benedict Wolff}
\IEEEauthorblockA{\textit{School of Electrical Engineering and Computer Science} \\
\textit{KTH Royal Institute of Technology}\\
Stockholm, Sweden \\
bjpwolff@kth.se \\
bjpw@live.de}
\and
\IEEEauthorblockN{Jacopo Bennati}
\IEEEauthorblockA{\textit{School of Electrical Engineering and Computer Science} \\
\textit{KTH Royal Institute of Technology}\\
Stockholm, Sweden \\
jbennati@kth.se \\
jacobbista.bennati@gmail.com}
}

\maketitle

\begin{abstract}

\copyrightnotice%
\Gls{LTM} is fundamental to \gls{LLM}-based agents in the emerging \gls{IoA}, where \glspl{DMAS} span cloud and edge networks. Existing evaluations are typically published by framework providers and focus on token usage and latency, rarely accounting for system-level cost or deployment in \gls{DMAS}. These gaps are addressed with an independent reproducible testbed that evaluates accuracy, latency, CPU time, peak RAM, disk I/O and network usage in a simulated cloud-edge environment. Three venture capital-funded frameworks spanning vector, graph, and hybrid architectures, namely mem0, Graphiti, and cognee, are compared alongside \gls{RAG} and full-context baselines on the LoCoMo benchmark under unconstrained and constrained network scenarios. Two clusters emerge: mem0, \gls{RAG}, and full-context reach 77\% to 81\% accuracy, while Graphiti and cognee reach only 55\% to 56\%, a gap driven by retrieval incompleteness rather than reasoning failure. The \gls{RAG} baseline matches the upper cluster at 8.4 times lower \gls{TCO} than mem0, and both are the only non-dominated backends on the Pareto frontier. Latency and bandwidth constraints as well as jitter leave retrieval quality unchanged for every backend, while vector-based \gls{LTM} incurs a modest latency penalty of 4\% to 5\% under edge-cloud constraints. Compression precision rather than context volume determines \gls{LTM} accuracy, as full-context forwarding underperforms mem0 despite supplying the entire conversation for each question.

\end{abstract}


\vspace{0.5cm}
\noindent\textbf{Open-source research:} the testbed, results, and notebooks for analysis and evaluation are available at: \url{https://github.com/wolffbe/dmas-memory}

\section{Introduction}
\label{sec:introduction}
\glsresetall

\Gls{LLM}-based agents are increasingly capable of decomposing goals into executable steps using in-context learning, integrating external tools and data sources as well as persisting thoughts using \gls{LTM}~\cite{zhang_survey_2025}. \par

\gls{LTM} refers to the ability of \gls{LLM}-based agents to store, organize, and retrieve information across interactions, enabling persistent context, behavioral adaptation, and personalization beyond single-turn reasoning~\cite{zhang_conversational_2025}. \Gls{RAG} is foundational to \gls{LTM} by embedding overlapping chunks of data into high-dimensional vector representations~\cite{lewis_retrieval-augmented_2020}. However, this static approach struggles with dynamic knowledge integration, introduces noise, and overlooks structural relationships. \gls{LTM} extends this paradigm into dynamic, agent-specific memory that accumulates over time, organized through textual, vector-based, graph-based and hybrid architectures, and supported by operations for data and knowledge indexing, retrieval, updating, and consolidation. \par

To evaluate these capabilities, \gls{LTM} benchmarks such as LoCoMo and LongMemEval have been proposed, focusing on tasks including multi-session reasoning, information retrieval, temporal consistency, and knowledge updates~\cite{maharana_evaluating_2024, wu_longmemeval_2025}. \par

Despite advances in \gls{LTM}, single-agent systems remain limited in addressing large-scale problems requiring distributed reasoning and coordination. This has led to the development of \glspl{MAS}, where specialized agents interact within a shared environment to enable collaborative decision-making, task delegation, and parallel problem solving~\cite{li_survey_2024}. \par

\Glspl{DMAS} further reduce reliance on centralized orchestration by enabling \gls{MAS} to coordinate across peer-to-peer networks, improving scalability, robustness, and privacy while avoiding bottlenecks associated with central controllers~\cite{chagas_distributed_2025, rojas-contreras_architectural_2026}. The vision is an \gls{IoA} that supports networks of heterogeneous agents processing information through agent integration protocols, messaging architectures, and dynamic teaming, while maintaining local data sovereignty, and being applied in long-term knowledge work and scientific research~\cite{chen_internet_2025, aminiranjbar_dawn_2025, wang_internet_2026}. \par

Despite these developments, three gaps remain. First, existing evaluations of \gls{LTM} are often not independent, as many comparisons are conducted or reported by the providers themselves. This introduces potential evaluation bias, since methods are often benchmarked against selective baselines under non-uniform conditions. Second, most studies focus on token-level performance metrics, while neglecting costs such as computation, network communication, and storage overhead introduced by \gls{LTM}. Third, the performance of \gls{LTM} in \gls{DMAS} remains underexplored, particularly in cloud-edge scenarios. \par

Filling these gaps is essential for further research on the \gls{IoA}. The performance of \gls{LTM} frameworks that agents in such a network build upon should be verifiable without any risk of evaluation bias. Given the wide range of technical constraints in network systems, understanding the resource consumption of \gls{LTM} in different network scenarios is of equal importance. \par

For that reason, this research aims to contribute to the fields of \gls{LTM}, \gls{DMAS} and the \gls{IoA} by answering the following three \glspl{RQ}: \par

\begin{itemize} 
    \item \textbf{RQ1:} \label{rq:1} What are the essential components of a uniform and independent testbed that enables reproducible comparison of different \gls{LTM} frameworks?
    \item \textbf{RQ2:} \label{rq:2} How do the cost and accuracy of \gls{LTM} frameworks perform in a \gls{DMAS} compared to a \gls{MAS}?
    \item \textbf{RQ3:} \label{rq:3} Which type of \gls{LTM} framework offers the best trade-off between cost and accuracy in a \gls{DMAS}?
\end{itemize} \par

\noindent In response to these \glspl{RQ}, this paper's main contributions are:

\begin{itemize}
    \item a testbed for evaluating \gls{LTM} frameworks in a \gls{DMAS}, measuring the accuracy and latency of responses as well as system-level costs in a simulated cloud-edge environment,
    \item independent benchmarks of three \gls{LTM} frameworks funded by venture capital, namely mem0 (vector-based), Graphiti (graph-based), and Cognee (hybrid), including full-context and \gls{RAG} baselines, using LoCoMo as the source of long-term conversational data, and 
    \item evaluations under unconstrained and constrained network scenarios, capturing the trade-off between \gls{LTM} accuracy, CPU time, peak RAM, disk I/O and network usage.
\end{itemize} \par

The rest of this paper is structured as follows. Section~\ref{sec:related-work} reviews the related work. Section~\ref{sec:method} presents the methodological approach to implementing and operating the testbed as well as analyzing its results described in Section~\ref{sec:results}. Section~\ref{sec:discussion} contains the discussion, limitations, and future work. \par

\section{Related Work}
\label{sec:related-work}

This section reviews prior work on \Gls{LLM}-based agents from three complementary perspectives. \gls{LTM} frameworks enable agents to memorize, organize, and retrieve knowledge across interactions. \gls{LTM} benchmarks evaluate these frameworks against standardized tasks. The \gls{IoA} addresses how agents interact and coordinate across peer-to-peer network infrastructures such as cloud-edge systems, extending \gls{MAS} to \gls{DMAS}. \par

\subsection*{Long-term Memory Frameworks}

One of the earliest vector-based implementations of \gls{LTM} is MemGPT, which augments the \gls{LLM} with a hierarchical memory architecture inspired by virtual memory. It partitions the fixed-size context window into system instructions, writable working memory, and a FIFO conversation buffer with recursive summarization, while storing extracted information in an external vector-indexed memory that is retrieved through function calls when additional context is required~\cite{packer_memgpt_2024}.

Mem0 extracts salient facts via \gls{LLM} calls against a rolling conversation summary, then reconciles them with semantically similar entries in a vector store through tool-call operations that add, update, delete, or skip memories~\cite{chhikara_mem0_2025}. The mem0\textsuperscript{g} variant additionally encodes these facts as a directed labeled graph of entities and typed relations. \par

Among other graph-based systems, Zep builds upon Graphiti, a temporally-aware Neo4j engine that ingests messages as episodes, extracts entities and inter-entity facts as labeled edges via \gls{LLM} prompts, and tracks a bitemporal validity window so that contradictory edges are invalidated rather than overwritten, while label-propagation clustering aggregates entities into a higher-level community subgraph~\cite{rasmussen_zep_2025}. Hindsight maintains four logical graphs: facts, experiences, entities, and beliefs, governed by retain, recall, and reflect operations~\cite{latimer_hindsight_2025}. \par

Hybrid systems combine vector and graph primitives. Cognee combines a Neo4j graph with a Qdrant vector store through a modular Extract--Cognify--Load pipeline that chunks documents and applies schema-constrained \gls{LLM} extraction to populate both stores with entities, relations, and summaries~\cite{markovic_optimizing_2025}. On the retrieval side, it exposes configurable strategies ranging from pure vector lookup over chunks to graph-completion routines that combine triplet retrieval with neighborhood traversal. Memoria pairs session-level summarization with a weighted knowledge graph that incrementally captures user traits, and behavioral patterns as structured entities and relationships~\cite{sarin_memoria_2025}. \par

These memory frameworks are almost universally benchmarked by their own authors, and the few cross-system comparisons among them run on heterogeneous harnesses, judges, and prompts, reducing evaluation to token consumption and task success. While latency is reported by several frameworks, CPU, RAM, disk I/O and network usage are rarely tracked. \par

\subsection*{Long-term Memory Benchmarks}

\Gls{LLM}-based agent evaluation is organized along two axes~\cite{mohammadi_evaluation_2025}. Evaluation objectives cover agent behavior, capabilities, reliability, safety, and alignment. The evaluation process consists of interaction mode, evaluation data, metrics computation methods, evaluation tooling, and evaluation contexts. Evaluation data is further split into datasets, benchmarks, and domain specificity. \par 

Latency, token usage, and cost are categorized as agent behaviour within evaluation objectives. Various benchmarks in this category show that evaluating the system-level cost and latency of agents is feasible. However, they are limited to GUIs or mobile agents rather than \gls{LTM}, \gls{MAS} or \gls{DMAS}. \par

Memory and context retention within agent capabilities is further refined into memory spans and memory forms~\cite{guan_evaluating_2026}. Memory spans include turn memory, conversation memory, and permanent memory, while memory forms cover textual form, such as complete, recent, retrieved and external interaction, and parametric form with fine-tuned memory and memory editing. \par

LoCoMo provides ten multi-session dialogues averaging 600 turns and 16,000 tokens over up to 32 sessions, with question categories spanning single-hop, multi-hop, temporal, open-domain, and adversary~\cite{maharana_evaluating_2024}. LongMemEval formalizes the five abilities information extraction, multi-session reasoning, temporal reasoning, knowledge updates, and abstention~\cite{wu_longmemeval_2025}. ES-MemEval extends this to emotional support, adding conflict detection and user modelling~\cite{chen_es-memeval_2026}. MS-TOD addresses multi-session task-oriented dialogue with 132 personas and intent-aligned memory annotations~\cite{du_memguide_2026} \par

Existing \gls{LTM} benchmarks primarily evaluate memory accuracy and rarely cover system-level costs. Cost-aware agent benchmarks do not cover \gls{LTM}, and neither line of work considers \gls{MAS} or \gls{DMAS}. To the best of our knowledge, this paper is the first to jointly evaluate \gls{LTM} accuracy and system-level cost of a \gls{DMAS} in a simulated cloud-edge scenario. \par

\subsection*{Internet of Agents}

Active fields of research on the \gls{IoA} include agent discovery, identity management, architectures, and edge deployments. \par

For agent discovery, a two-stage framework has been proposed in which agents first publish machine-interpretable capability descriptions and then perform context-aware search, ranking, and composition to assemble suitable agents for a given task~\cite{guo_agent_2026}. The scheme integrates capability modelling, scalable and updatable indexing, and continual discovery. \par

Regarding identity management, \gls{LLM}-based agents have been integrated with self-sovereign identity to mitigate prompt injection, unauthorized access, and API exploitation~\cite{aydeger_decentralized_2026}. \par

With respect to agent architectures, a \gls{DMAS} underpinned by a blockchain and a trust-aware communication protocol has been proposed, providing verifiable interaction cycles, non-repudiation, and conditional confidentiality~\cite{ding_decentralized_2025}. \par

In the field of edge deployments, several lines of work address the gap between resource-constrained devices and the demands of \gls{LLM}-based agents. CASK reduces inference-time memory in \gls{MAS} through a dynamic sparse attention module and adaptive \gls{KV}-cache compression, cutting memory usage by up to 40\% while retaining over 95\% of baseline accuracy on LongBench~\cite{mohammed_context_2025}. An adaptive placement and migration framework uses ant colony optimization and \gls{LLM}-based decision making on top of AgentScope to reduce deployment latency and migration cost in dynamic edge environments~\cite{wang_adaptive_2025}. Closest to the present work, a heterogeneous edge testbed on K3s, a lightweight Kubernetes distribution, with eight Jetson Nano and Raspberry Pi clients has been constructed, simulating realistic 4G/5G network conditions through Linux Traffic Control~\cite{tadi_performance_2025}. Quantization, asynchronous federated learning, distributed caching, and network-aware scheduling are evaluated on Llama 3 8B, with a reported 75\% bandwidth reduction, 45\% latency improvement, and 61\% energy decrease per inference. The testbed methodology is closely related, but the focus lies on federated training and inference rather than evaluating \gls{LTM} frameworks. \gls{LTM} beyond the \gls{KV}-cache is not considered. \par

Across these directions, no work to date evaluates \gls{LTM} frameworks in \gls{DMAS} or the \gls{IoA}. Discovery, identity, and architecture work treat \gls{LTM} as an internal property of individual agents. Edge-deployment work either compresses the \gls{KV}-cache at inference time or focuses on training-time federated workloads. The testbed proposed in this paper fills this gap by evaluating accuracy and system-level cost of \gls{LTM} frameworks under simulated conditions that resemble \gls{DMAS} deployments. \par

\section{Method}
\label{sec:method}

The methodology applied in this study follows the experimental framework described by Wohlin et al.~\cite{wohlin_experimentation_2024}, which structures empirical software engineering experiments into five phases, namely scoping, planning, operation, analysis and interpretation as well as presentation and packaging. The first four phases are described in this section. Presentation and packaging is realized through sections \ref{sec:results} and \ref{sec:discussion}. \par

\subsection*{Scoping the Experiments}

Scoping consists of defining the object of study, purpose, quality focus, perspective and context. The goal of the experiment is formulated using a structured Goal-Question-Metric template covering the object of study, purpose, quality focus, perspective, and context~\cite{basili_goal_1994}. \par

\subsubsection*{Object of Study} The objects of study are three \gls{LTM} frameworks, all implemented as Python-based Docker services and selected to span the architectural spectrum from vector to graph storage: mem0, a vector-based system with \gls{LLM}-driven compression that raised \$24 million USD~\cite{singh_mem0_2025}; Graphiti, a graph engine built on Neo4j released as the open-source core of Zep, which has raised \$2.3 million USD in pre-seed funding~\cite{crunchbase_inc_zep_2026}; and cognee, a hybrid framework combining vector and graph architectures, which has raised \$7.5 million USD~\cite{markovic_cognee_2026}. The candidate pool was restricted to \gls{LTM} with both venture capital funding and published benchmarks, a set that also includes Hindsight and Letta, formerly known as MemGPT~\cite{latimer_hindsight_2025, packer_memgpt_2024}. From this pool, mem0, cognee, and Graphiti were selected for their comparable scale and intersecting technical stack consisting of Python, Docker, Neo4j and Qdrant. Hindsight and Letta were excluded as both market themselves as entire platforms for \gls{LLM}-based agents rather than exclusively \gls{LTM}. \gls{RAG} and full-context implementations, where an \gls{LLM} is given an entire conversation with each question, serve as baselines. \par

\subsubsection*{Purpose} The purpose of the experiment is to compare these frameworks under two consistent network scenarios, characterizing the trade-off between memorizing and retrieving knowledge as well as system-level resource consumption. \par

\subsubsection*{Quality Focus} Two quality attributes are considered. The first is memory accuracy, capturing whether an agent can correctly answer questions that depend on \gls{LTM}. The second is system-level cost, capturing \gls{LLM} usage in tokens and USD, CPU time, peak RAM, disk I/O and network usage. \par

\subsubsection*{Perspective} The experiment is conducted from the perspective of  researchers developing infrastructure for the \gls{IoA}, where independence from any commercial interest and applicability to \gls{DMAS} and the \gls{IoA} are central concerns. \par

\subsubsection*{Context} The experiment takes place on a simulated cloud-edge testbed in which agents communicate across a virtual network link whose bandwidth, latency, and jitter is programmatically fixed to reproduce realistic conditions. \par

\subsection*{Planning the Experiments}

The planning phase covers selecting the context, formulating hypothesis, defining variables, designing the experiment, choosing the right instrumentation, and mitigating four classes of threats to validity. \par

\subsubsection*{Context Selection} The experiment is technology-oriented, operating on \gls{LLM}-based agents rather than human subjects, is simulated and conducted offline\footnote{In Wohlin's framework, \emph{offline} refers to a controlled experimental setting as opposed to a live industrial deployment.}. The benchmark comprises ten conversations replayed once in each framework and network condition. \gls{LLM} temperature is fixed at zero and network shaping is controlled, which minimizes within-cell variance. \par

\subsubsection*{Hypothesis formulation} For \hyperref[rq:2]{\gls{RQ}2}, the null hypothesis \label{h:02} $H_0^{(2)}$ states that the accuracy and cost of the \gls{LTM} frameworks are not affected by the transition between network scenarios. The alternative hypothesis \label{h:12} $H_1^{(2)}$ states that at least one framework exhibits a change in accuracy or cost. For \hyperref[rq:3]{\gls{RQ}3}, the null hypothesis \label{h:03} $H_0^{(3)}$ states that no significant difference in the accuracy-cost trade-off exists between mem0, Graphiti and cognee. The alternative hypothesis \label{h:13} $H_1^{(3)}$ states that at least one framework exhibits a significantly different trade-off. \par

\subsubsection*{Variables} Two independent variables are manipulated. The first is the \gls{LTM} framework, with five levels, namely the three frameworks mem0, Graphiti, and cognee, and the two baselines \gls{RAG} and full-context. The second is the network scenario, with two levels. The unconstrained scenario adds no artificial latency, jitter or bandwidth restrictions between edge and cloud components. The constrained scenario models a edge-to-cloud backhaul link, with 150\,ms of latency, 512\,KB/s of bandwidth, and 30\,ms of jitter, representative of embedded edge devices accessing cloud services over a wide-area network~\cite{3rd_generation_partnership_project_3rd_2026}. \par

The dependent variables fall into four groups. Latency consists of wall-clock time, isolated compute time, and any flush time incurred while persisting state to the memory backend. Resource utilization, recorded separately for the edge and cloud components, comprises CPU time, peak RAM, disk I/O, and network I/O. \gls{LLM} usage and cost is logged as token counts and USD cost for edge-side and cloud-side calls, with the responder's context window consumption tracked as a separate line item. Answer quality, the fourth group, pairs each answer with its gold answer, records the number of memories returned by retrieval and the number of memory accesses, and is evaluated using an \gls{LLM}-as-a-judge approach~\cite{gu_survey_2026}. \par

For each question, the judge is given the question, the gold answer from the benchmark, and the produced answer. It then returns a verdict based on the same prompt that was used to evaluate Graphiti due to its restrictiveness compared to the prompt used in mem0's benchmark~\cite{rasmussen_zep-paperskg_architecture_agent_memorylocomo_eval_2025, rathi_memory-benchmarksbenchmarkslocomopromptspy_2026}. To mitigate sampling variance, the judge is invoked three times per question and the final verdict is taken by majority vote. \par

\subsubsection*{Experiment design} A full factorial design is adopted, combining each \gls{LTM} framework with each network scenario, yielding ten experimental treatments. All ten conversations of the LoCoMo dataset are processed within each treatment. \par

\subsubsection*{Instrumentation}

\begin{figure}[!t]
    \centering
    \includegraphics[width=0.85\columnwidth]
    {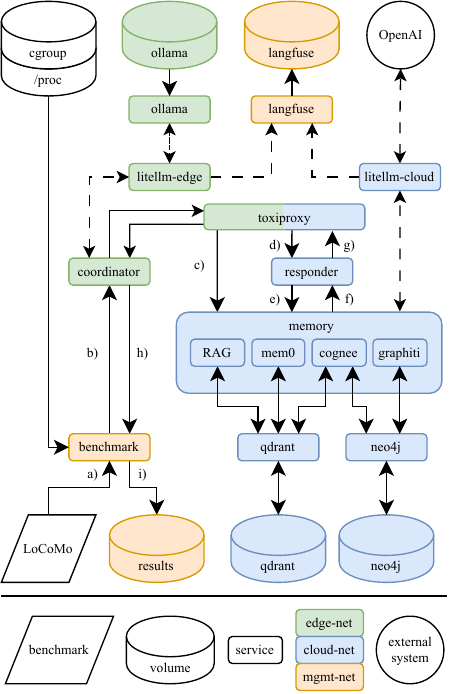}
    \caption{Testbed architecture including the benchmark, coordinator, responder and memory. \gls{LLM} invocations to LiteLLM, Langfuse and OpenAI are marked as dashed lines. Components are split across an edge, cloud and management network. Toxiproxy is assigned to the edge and cloud networks.}
    \label{fig:sys_arch}
\end{figure}

The testbed shown in Figure~\ref{fig:sys_arch} simulates the \gls{DMAS} on a single host using Docker Compose, with the edge--cloud scenarios implemented through three isolated Docker bridge networks rather than physical hosts. The \texttt{edge-net} hosts the \texttt{coordinator} and the \gls{LLM} runtime \texttt{ollama}~\cite{ollama_ollamaollama_2026} for local \glspl{SLM}. The \texttt{cloud-net} hosts the \texttt{memory} and \texttt{responder} services together with the two databases used by the \gls{LTM} frameworks, namely \texttt{qdrant} used by mem0, cognee and the RAG baseline, and \texttt{neo4j} used by Graphiti and cognee. The \texttt{mgmt-net} bridge isolates the \texttt{benchmark} service and the \texttt{langfuse} observability stack~\cite{langfuse_gmbh_langfuselangfuse_2026} so that control-plane calls and metric scrapes never traverse the measured data plane. The role of Langfuse is to capture traces throughout each experiment. All \gls{LLM} traffic is routed through the AI gateway LiteLLM~\cite{litellm_berriailitellm_2026}, deployed as two instances reachable only from their own subnet, with \texttt{litellm-edge} transmitting to the local \texttt{ollama} instance and \texttt{litellm-cloud} dispatching to the OpenAI platform. LiteLLM is responsible for capturing the token usage, and converting it to USD. The only container in both data-plane subnets is \texttt{toxiproxy}~\cite{shopify_toxiproxy_2025}, the gateway between \texttt{edge-net} and \texttt{cloud-net} that applies latency, jitter, and bandwidth limits to all edge--cloud traffic. While \texttt{proc} is used to track network traffic, \texttt{cgroup} captures CPU, RAM and disk I/O. \par

As indicated through the letters (a-g) in Figure~\ref{fig:sys_arch}, \texttt{benchmark} loads the LoCoMo benchmark from GitHub~(a), decomposes each conversation into non-overlapping chunks of one turn, where one turn corresponds to a single message in the conversation, and forwards them sequentially to the \texttt{coordinator}~(b), tagging every request with the target memory framework. The experiment proceeds in two phases. During the loading phase, the \texttt{coordinator} forwards each turn across the edge--cloud boundary through \texttt{toxiproxy} to the \texttt{memory} agent's \texttt{/remember} endpoint~(c), which routes the request to the active backend among RAG, mem0, cognee, and Graphiti. During the Q\&A phase, the \texttt{benchmark} issues questions through the \texttt{coordinator}, which forwards each question across the same boundary to the \texttt{responder}'s \texttt{/ask} endpoint~(d). The \texttt{responder} retrieves a fixed top $k=20$ related memories from the \texttt{memory} agent over the intra-cloud path~(e,~f). The \texttt{responder} may search for memories in parallel up to four times. After retrieving all relevant memories, it generates the answer, which is returned through \texttt{toxiproxy}~(g) and the \texttt{coordinator}~(h) back to the \texttt{benchmark}. Finally, the \texttt{benchmark} appends all questions, answers, and metrics to a CSV file on disk~(i). \par

The \gls{SLM} \textit{qwen2.5:3b-instruct-q4\_K\_M} using three billion parameters optimized for tool calling, drives the Q\&A phase through function calls and is bypassed during the loading phase. The cloud \gls{LLM} \textit{gpt-4o-mini} is shared by every cloud-side call, namely the \texttt{responder}'s answer generation and the load-phase extraction performed by mem0, Graphiti, and cognee. Vector embeddings for mem0, Graphiti, cognee, and the \gls{RAG} baseline are produced by OpenAI's \textit{text-embedding-3-small}. \par

\subsubsection*{Threats to Validity} 

Following Wohlin's four-dimensional model~\cite{wohlin_experimentation_2024}, four classes of threats are considered. \par

For internal validity, all components involved are used at fixed versions. Memory backends are logically wiped between conversations while the testbed stays up, preserving the containers, networks, the \texttt{ollama} model cache, and the \texttt{neo4j} schema layer of indexes and uniqueness constraints.  \par

During a dedicated warmup phase, an explicit initial call to the memory framework is issued to absorb any lazy initialization and first-invocation client overhead. Its resource counters and latency are excluded from the load and Q\&A phases that follow. The \texttt{toxiproxy} toxics are set at the beginning of each conversation to match the configured network profile. To keep asynchronous work inside each call's measurement window, \texttt{neo4j}'s \texttt{db.awaitIndexes} is invoked synchronously inside every \texttt{/memorize} handler so index population completes before the response returns, and the \texttt{benchmark} service holds its post-call \texttt{cgroup} snapshot until cloud-side disk I/O quiesces. This post-response wait is recorded separately from the actual computation as flush. \par

To minimize the risk that performance differences reflect researcher-introduced implementation choices rather than the frameworks themselves, the testbed code for mem0 and Graphiti is adapted from their authors' own repositories~\cite{kumar_mem0aimemory-benchmarks_2026, rasmussen_getzepzep-papers_2026}. A residual threat is that these benchmarks have been mutually disputed by competing framework providers, namely Zep has documented choices in mem0's published evaluation that may understate Graphiti's performance~\cite{chalef_is_2025} and mem0 has raised methodological concerns about Zep's evaluation~\cite{yadav_revisiting_2025}. \par

External validity is bounded by the choice of dataset, the underlying \glspl{LLM}, and the fact that the cloud--edge separation is simulated on a single machine rather than physically distributed across two hosts. To probe whether observed differences depend on the specific framework set, the \gls{RAG} and full-context baselines are included. \par

Construct validity is addressed by operationalizing answer correctness through an \gls{LLM}-as-judge with majority voting over three independent calls, which mitigates mono-method bias in the verdict, and by operationalizing system-level cost as a multivariate vector of CPU, RAM, disk I/O, network, and token spend rather than a single proxy, so that no backend's strengths can hide on an unmeasured axis. To further separate retrieval failure from reasoning error within the outcome variable, the \texttt{responder}'s system prompt instructs to not reply when the retrieved context is insufficient rather than to guess, so that absent memory is recorded as \textit{unknown} instead of being miscoded as \textit{wrong}. A residual threat is that the judge model shares a family with the responder, which may inflate agreement through self-enhancement bias. Majority voting attenuates but does not eliminate this effect~\cite{zheng_judging_2023}. \par

Conclusion validity is based on 1,540 question–answer pairs across all ten LoCoMo benchmark conversations, replayed under both network scenarios and covering all questions in categories one through four. Following mem0 and zep, category five questions, which have reference answers of the form ``not mentioned'', are excluded as they conflate retrieval performance with the refusal behaviour of the responder \gls{LLM}, rather than isolating the performance of the memory backend itself. \par

\subsection*{Operation}

\subsubsection*{Preparation} The testbed is deployed on AWS EC2 \texttt{c7i.4xlarge} instances with 16 vCPUs, 32\,GiB RAM, a 4th-generation Intel Xeon Scalable processor, and a 120\,GB EBS gp3 volume, running Ubuntu Server 26.04 LTS. \par

\subsubsection*{Execution} The experiment runs in parallel across four EC2 instances, one per framework setup (cognee, mem0, graphiti as well as RAG and full-context), each executing both network scenarios and writing its results to CSVs on disk. \par

\subsubsection*{Data validation} Each CSV is checked for the expected number of rows, missing fields, and anomalous resource readings. The validated dataset forms the input to the analysis in the next section, where the methodology applied to evaluate the results is described. \par

\subsection*{Analysis and Interpretation}

Analysis and interpretation covers the statistical tests applied to the outcomes, the pricing model used to aggregate resource consumption into the \gls{TCO}, and the dominance rule that summarizes the cost--accuracy trade-off. \par

The cost and accuracy for each backend are compared pairwise using the paired Wilcoxon signed-rank test~\cite{wilcoxon_individual_1945}, with the ten conversations from LoCoMo forming the pairing unit. A Bonferroni correction across the ten framework pairs sets the per-test significance threshold to $\alpha=0.005$~\cite{bonferroni_teoria_1936}. The non-parametric form is preferred to a paired $t$-test because normality of the per-conversation distributions cannot be assumed at $n=10$. The same procedure pairs the constrained and unconstrained observations of each conversation in order to test for network sensitivity within each \gls{LTM}. \par

\Gls{TCO} is computed as the sum of edge and cloud resource consumption per \gls{LTM} framework across both the loading and Q\&A phases, priced at the AWS Fargate \texttt{us-east-1} on-demand rates listed in Table~\ref{tab:tco}. Cloud-to-edge egress traffic between the \texttt{coordinator} and the \texttt{responder} would cross the public internet in a real deployment and is therefore charged at \$0.09 per GB. Edge-to-cloud ingress is charged at \$0.00 per GB. Traffic within \texttt{cloud-net} as well as egress traffic from \texttt{litellm-cloud} to OpenAI are also treated at zero cost. \par

The cost--accuracy trade-off is summarized as a statistical Pareto frontier over the five backends. For each ordered pair of backends $(A, B)$, $A$ is said to dominate $B$ if its \gls{TCO} is strictly lower than $B$'s and its accuracy is either statistically equivalent to or higher than $B$'s, where statistical equivalence is taken as a failure to reject equality under the same paired Wilcoxon signed-rank test introduced above at $\alpha=0.005$. The frontier is then defined as the set of backends not dominated by any other under this rule. The statistical relaxation of the standard dominance criterion prevents a marginal accuracy difference that the data cannot resolve from disqualifying a cheaper backend, and prevents a cheaper backend with significantly lower accuracy from being placed on the frontier. \par

\begin{table}[!t]
\centering
\caption{\gls{AWS} and OpenAI resources, services and rates used for calculating the \gls{TCO} of each \gls{LTM} and baseline}
\label{tab:tco}
\begin{tabular}{@{}llr@{}}
\toprule
\textbf{Resource} & \textbf{Service} & \textbf{Rate} \\
\midrule
Disk                & \gls{AWS} Fargate (20\,GB free)   & \$0.000109/GB-hr \\
RAM                 & \gls{AWS} Fargate (Linux/x86)     & \$0.004445/GB-hr \\
vCPU time           & \gls{AWS} Fargate (Linux/x86)     & \$0.04048/vCPU-hr \\
\midrule
Cloud-to-edge egress & Amazon EC2 Data Transfer   & \$0.09/GB \\
Edge-to-cloud ingress & Amazon EC2 Data Transfer  & \$0.00/GB \\
Intra-cloud network  & Amazon EC2 Data Transfer   & \$0.00/GB \\
\midrule
LLM input           & OpenAI gpt-4o-mini          & \$0.15/M tokens \\
LLM cached input    & OpenAI gpt-4o-mini          & \$0.075/M tokens \\
LLM output          & OpenAI gpt-4o-mini          & \$0.60/M tokens \\
\bottomrule
\end{tabular}

\smallskip
{\footnotesize Sources: AWS Fargate~\cite{amazon_web_services_inc_aws_2026};
Amazon EC2 Data Transfer~\cite{amazon_web_services_inc_ec2_2026};
OpenAI~\cite{openai_developers_pricing_2026}.}
\end{table}

\section{Results}
\label{sec:results}

Two anomalies occur in the data. Firstly, the \texttt{responder} queried \gls{LTM} multiple times for only 0.83\% of questions across all conversations, frameworks, and network scenarios. To reduce ambiguity between experimental treatments, these questions are dropped across all \gls{LTM} frameworks and baselines. Secondly, despite best efforts to eliminate this error through automated pruning, the \glspl{LLM} context window was exceeded for 0.03\% of all questions while running full-context. These questions are also removed from the results. \par

The cost of the warm-up and flush phases are not described further, as their impact does not further influence the results. Likewise, the paired Wilcoxon signed-rank test indicates no statistically significant difference between the unconstrained and constrained network scenarios (all $p{>}0.07$). Consequently, the unconstrained scenario is not considered in detail, since its exclusion does not affect the validity of the analysis. \par

The absence of statistically significant differences under artificially introduced latency, bandwidth constraints, and jitter further strengthens the robustness of the data from the constrained scenario, as it indicates that the reported results remain stable across varying network properties. While this limits differentiation between network scenarios, it yields repeated measurements under equivalent treatments, increasing the sample size and improving the reliability and robustness of the observed results across multiple experimental iterations. \par

The results are presented in four parts. The first part covers the loading phase, the second part the Q\&A phase, the third part aggregates the cost of both phases into the \gls{TCO} of each \gls{LTM}, and the fourth parts calculates the cost-accuracy analysis based on the accuracy and \gls{TCO}. \par

\begin{table*}[t]
\centering
\caption{Latency and system-level cost of three \gls{LTM} frameworks and two baselines in the constrained cloud-edge network scenario. The RAM peak subtotals and total reflect the highest peak identified during experimentation.}
\label{tab:cost}
\begin{tabular}{ll r rrrrr rrrr}
\toprule
 & & & \multicolumn{5}{c}{\textbf{Cloud}} & \multicolumn{4}{c}{\textbf{Edge}} \\
\cmidrule(lr){4-8} \cmidrule(lr){9-12}
 & & \textbf{Latency} & \textbf{CPU} & \textbf{RAM peak} & \textbf{Disk I/O} & \textbf{Network} & \textbf{OpenAI} & \textbf{CPU} & \textbf{RAM peak} & \textbf{Disk I/O} & \textbf{Network} \\
\textbf{Phase} & \textbf{Memory} & (hours) & (hours) & (MB) & (MB) & (MB) & (USD) & (hours) & (MB) & (MB) & (MB) \\
\midrule
\multirow{6}{*}{Loading} & cognee & 8.44 & 4.36 & 962.97 & 21,664.74 & 5.48 & \$1.32 & 0.09 & 2,562.79 & 0.92 & 7.69 \\
 & full-context & 0.61 & 0.02 & 18.39 & 1.23 & 5.27 & \$0.00 & 0.01 & 2,343.19 & 0.11 & 7.80 \\
 & Graphiti & 9.69 & 0.56 & 866.93 & 7,904.48 & 5.39 & \$5.49 & 0.11 & 2,591.10 & 0.01 & 7.65 \\
 & mem0 & 4.44 & 0.26 & 126.50 & 2,156.02 & 5.45 & \$4.82 & 0.06 & 2,592.46 & 0.00 & 7.63 \\
 & RAG & 0.93 & 0.08 & 70.01 & 993.21 & 5.48 & \$0.01 & 0.02 & 2,342.22 & 0.00 & 7.62 \\
\cmidrule(l){2-12}
 & \textit{Subtotal} & \textit{24.11} & \textit{5.28} & \textit{962.97} & \textit{32,719.68} & \textit{27.07} & \textit{\$11.64} & \textit{0.29} & \textit{2,592.46} & \textit{1.04} & \textit{38.39} \\
\midrule
\multirow{6}{*}{Q\&A} & cognee & 2.93 & 0.80 & 1,001.63 & 176.07 & 1.22 & \$1.05 & 5.20 & 2,841.20 & 0.00 & 1.58 \\
 & full-context & 3.89 & 0.09 & 18.96 & 3.81 & 1.41 & \$10.01 & 5.08 & 2,375.77 & 0.00 & 1.61 \\
 & Graphiti & 2.76 & 0.27 & 876.25 & 17.46 & 1.27 & \$1.00 & 5.25 & 2,856.97 & 0.07 & 1.62 \\
 & mem0 & 2.04 & 0.07 & 132.02 & 4.23 & 1.39 & \$0.27 & 5.75 & 2,786.05 & 0.01 & 1.60 \\
 & RAG & 1.90 & 0.06 & 79.76 & 5.08 & 1.34 & \$0.40 & 5.18 & 2,501.20 & 0.00 & 1.60 \\
\cmidrule(l){2-12}
 & \textit{Subtotal} & \textit{13.52} & \textit{1.29} & \textit{1,001.63} & \textit{206.65} & \textit{6.63} & \textit{\$12.73} & \textit{26.46} & \textit{2,856.97} & \textit{0.08} & \textit{8.01} \\
\midrule
\multicolumn{2}{l}{Total} & 37.63 & 6.57 & 1,001.63 & 32,926.33 & 33.70 & \$24.37 & 26.75 & 2,856.97 & 1.12 & 46.40 \\
\bottomrule
\end{tabular}
\end{table*}

\subsection*{Loading Long-Term Memory}

As shown in Table~\ref{tab:cost}, loading all ten conversations from LoCoMo required 24.11 hours, 32,719.68 MB of disk I/O in the cloud, 1.04 MB of disk I/O at the edge, 38.39 MB of edge-to-cloud traffic, and 27.07 MB of cloud-to-edge traffic. CPU time totaled 0.29 hours at the edge and 5.28 hours in the cloud. RAM usage peaked at 2,592.46 MB at the edge and 962.97 MB in the cloud. The OpenAI cost was \$11.64 USD. \par

Full-context was the fastest baseline, requiring only 0.61 hours to load all memories, followed by \gls{RAG} at 0.93 hours, mem0 at 4.44 hours, cognee at 8.44 hours, and Graphiti at 9.69 hours. Disk I/O was lowest for full-context at 1.23 MB, followed by \gls{RAG} at 993.21 MB, mem0 at 2,156.02 MB, Graphiti at 7,904.48 MB, and cognee at 21,664.74 MB. Network traffic between the \texttt{coordinator} and \texttt{responder} remained relatively stable across all frameworks, ranging from 13.08 MB for Graphiti to 13.28 MB for full-context when combining bidirectional traffic. CPU usage increased across systems, with full-context requiring the least total CPU time at 0.03 hours, followed by \gls{RAG} at 0.10 hours, mem0 at 0.32 hours, Graphiti at 0.67 hours, and cognee at 4.45 hours. Peak cloud RAM usage increased from full-context at 18.39 MB, to \gls{RAG} at 70.01 MB, mem0 at 126.50 MB, Graphiti at 866.93 MB, and cognee at 962.97 MB. OpenAI cost likewise increased from lowest to highest, with full-context at zero cost, followed by \gls{RAG} at \$0.01 USD, cognee at \$1.32 USD, mem0 at \$4.82 USD, and Graphiti at \$5.49 USD. \par

\subsection*{Answering Questions Using Long-Term Memory}

The outcomes of the Q\&A phase are further divided into two sections, namely cost and accuracy, capturing resource consumption and response correctness respectively. \par

\subsubsection*{Cost}

After excision of the erroneous questions, answering 7,626 questions across all frameworks, baselines, and conversations in LoCoMo required 13.52 hours, 206.65 MB of disk I/O in the cloud, 0.08 MB of edge disk I/O, 8.01 MB of edge-to-cloud traffic, and 6.63 MB of cloud-to-edge traffic. CPU time totaled 26.46 hours at the edge and 1.29 hours in the cloud. RAM usage peaked at 2,856.97 MB at the edge and 1,001.63 MB in the cloud. The total cost of OpenAI’s \gls{LLM} inference and embeddings was \$12.73 USD. \par

\gls{RAG} was the fastest baseline for answering all questions, requiring 1.90 hours, followed by mem0 at 2.04 hours, Graphiti at 2.76 hours, cognee at 2.93 hours, and full-context at 3.89 hours. Disk I/O was lowest for full-context at 3.81 MB, followed by mem0 at 4.23 MB, \gls{RAG} at 5.08 MB, Graphiti at 17.46 MB, and cognee at 176.07 MB. Network traffic was lowest for cognee at 2.80 MB, followed by Graphiti at 2.89 MB, \gls{RAG} at 2.94 MB, mem0 at 2.99 MB, and full-context at 3.02 MB. Total CPU usage increased across systems, with mem0 requiring the least at 5.82 hours, followed by full-context at 5.17 hours, \gls{RAG} at 5.24 hours, cognee at 6.00 hours, and Graphiti at 5.52 hours. Peak cloud RAM usage increased from full-context at 18.96 MB, to \gls{RAG} at 79.76 MB, mem0 at 132.02 MB, Graphiti at 876.25 MB, and cognee at 1,001.63 MB. OpenAI cost ranged from lowest to highest, with mem0 at \$0.27 USD, \gls{RAG} at \$0.40 USD, Graphiti at \$1.00 USD, cognee at \$1.05 USD, and full-context at \$10.01 USD. \par

\subsubsection*{Accuracy}

\begin{table}[!t]
\centering
\caption{Accurate, incorrect and unknown responses are computed per conversation and then averaged over all conversations. $\pm$~denotes the standard deviation across all conversations.}
\label{tab:retention_macro}
\begin{tabular}{l rrr}
\toprule
\textbf{Memory} & \textbf{Accurate (\%)} & \textbf{Incorrect (\%)} & \textbf{Unknown (\%)} \\
\midrule
cognee & 55.27 $\pm$ 3.60 & 16.55 $\pm$ 5.11 & 28.18 $\pm$ 4.92 \\
full-context & 77.16 $\pm$ 5.21 & 13.17 $\pm$ 3.50 & 9.67 $\pm$ 5.40 \\
Graphiti & 56.03 $\pm$ 5.56 & 18.39 $\pm$ 3.47 & 25.58 $\pm$ 3.38 \\
mem0 & 81.08 $\pm$ 3.19 & 10.54 $\pm$ 2.64 & 8.38 $\pm$ 2.14 \\
RAG & 78.31 $\pm$ 5.75 & 10.56 $\pm$ 3.61 & 11.14 $\pm$ 3.95 \\
\bottomrule
\end{tabular}
\end{table}

The accuracy of each \gls{LTM} is shown in Table~\ref{tab:retention_macro}. Mem0 attains the highest performance, correctly answering 81.08\% of questions, with 10.54\% answered incorrectly and 8.38\% marked as unknown. The \gls{RAG} baseline follows with 78.31\% correct answers, 10.56\% incorrect answers, and 11.14\% unknown responses. The full-context baseline reaches 77.16\% accuracy, with 13.17\% incorrect and 9.67\% unknown responses. Graphiti records 56.03\% correct answers, with 18.39\% incorrect and 25.58\% unknown responses. Cognee achieves 55.27\% correct, 16.55\% incorrect and 28.18\% unknown responses. \par

\begin{table}[!t]
\centering
\caption{Accuracy of three \gls{LTM} frameworks and two baselines across four LoCoMo question categories in the constrained cloud-edge network scenario. Results are descriptive due to the limited amount of questions per category.}
\label{tab:categories}
\begin{tabular}{lcccc}
\toprule
\textbf{Memory}
  & \textbf{Single-hop}
  & \textbf{Multi-hop}
  & \textbf{Temporal}
  & \textbf{Open-domain} \\
\midrule
cognee        & 52.79\% & 29.88\% & 43.14\% & 67.66\% \\
full-context  & 67.94\% & 62.84\% & 36.65\% & 89.75\% \\
Graphiti      & 56.00\% & 52.59\% & 31.83\% & 59.85\% \\
Mem0          & 71.09\% & 74.79\% & 55.48\% & 88.81\% \\
\gls{RAG}     & 72.12\% & 73.08\% & 45.32\% & 85.30\% \\
\bottomrule
\end{tabular}
\end{table}

The accuracy of each \gls{LTM} is further considered across four question categories as presented in Table~\ref{tab:categories}: single-hop and multi-hop depending on whether one or multiple retrieval steps are required, temporal questions depending on whether time-based reasoning is required, and open-domain questions depending on whether questions span broad topics or a restricted setting. Per-category accuracy is reported  as indicative trends only, since the per-conversation per-category sample size is too small to support statistical significance testing. Multi-hop questions show the largest variation between systems, ranging from 74.79\% for mem0 and 73.08\% for \gls{RAG} down to 29.88\% for cognee. Open-domain questions are the easiest overall, with no framework or baseline achieving less than 59.85\% accuracy. \par

\subsection*{Total Cost of Ownership}
    
Based on the unit prices provided in Table~\ref{tab:tco}, the \gls{TCO} reflects both system-level and \gls{LLM}-related expenditures aggregated across the loading and Q\&A phases, covering both cloud and edge resources. \gls{RAG} has the lowest \gls{TCO} at \$0.65 USD, followed by cognee at \$2.99 USD, mem0 at \$5.43 USD, Graphiti at \$6.95 USD, and full-context at \$10.22 USD. Given that disk usage remains below 20 GB throughout all experiments, storage-related costs are effectively zero under the applied pricing model. \par

\subsection*{Cost-Accuracy Trade-Off}

\begin{figure}[!t]
  \centering
  \includegraphics[width=\columnwidth]{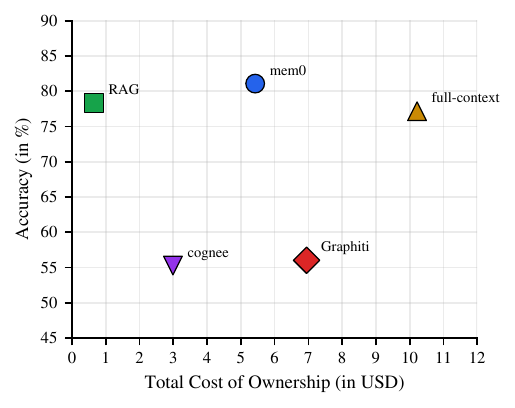}
  \caption{Cost-accuracy trade-off of three \gls{LTM} frameworks and two baselines in the constrained network scenario. RAG and mem0 are pareto-optimal.}
  \label{fig:pareto}
\end{figure}

Figure~\ref{fig:pareto} shows that only \gls{RAG} and mem0 lie on the Pareto frontier. \gls{RAG} attains a mean accuracy of 78.31\% at a \gls{TCO} of \$0.65 USD. Mem0 reaches 81.08\% at a \gls{TCO} of \$5.43 USD. The full-context baseline records 77.16\% at a \gls{TCO} of \$10.22 USD. Graphiti scores 56.03\% at a \gls{TCO} of \$6.95 USD, and cognee records 55.27\% at a \gls{TCO} of \$2.99 USD. \par

\section{Discussion}
\label{sec:discussion}

The following section interprets the empirical findings in relation to the three research questions and four hypotheses by examining the mechanisms underlying the observed differences in retrieval accuracy, latency sensitivity, and system-level cost across the evaluated \glspl{LTM} and baseline configurations. \par

\begin{figure}[!t]
  \centering
  \includegraphics[width=\columnwidth]{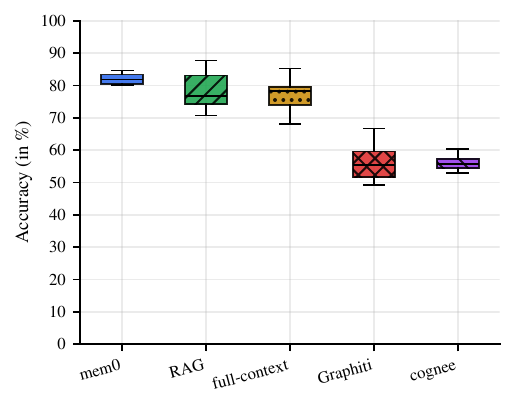}
  \caption{Accuracy of three \gls{LTM} frameworks and two baselines across ten LoCoMo conversations in the constrained cloud-edge network scenario. Two clusters emerge: the first one containing mem0, \gls{RAG} and full-context, and the second one containing graphiti and cognee.}
  \label{fig:accuracy}
\end{figure}

As shown in Figure~\ref{fig:accuracy}, accuracy separates into two statistically distinct clusters: Mem0, \gls{RAG}, and full-context span 77.16\% to 81.08\%, while Graphiti and cognee reach only 56.03\% and 55.27\% respectively. All cross-cluster comparisons are significant at $p = 0.002$ under a paired Wilcoxon signed-rank test with Bonferroni correction, and no within-cluster differences reach significance (mem0 versus \ \gls{RAG}: $p = 0.084$, Graphiti versus \ cognee: $p = 0.845$). The ranking is stable across all ten LoCoMo conversations. \par

The gap is driven by retrieval incompleteness rather than reasoning failure. As presented in Table~\ref{tab:retention_macro}, Graphiti and cognee return unknown responses on 25.58\% and 28.18\% of questions respectively, compared to 8.38\% to 11.14\% for the upper cluster, and a paired Wilcoxon test on unknown rates reproduces the same structure at $p = 0.002$. When graph-based and hybrid \gls{LTM} do retrieve context, their incorrect answer rates are only moderately higher than those of the upper cluster. The dominant failure mode is therefore the absence of retrieved content, not its misinterpretation. Graph-based ingestion introduces entity resolution, relation labeling, and schema-constrained extraction, each an opportunity for information loss before retrieval is attempted. This compounds on multi-hop questions, where the cross-cluster gap is largest, ranging from 74.79\% for mem0 down to 29.88\% for cognee. \par

Counterintuitively, full-context scores below mem0, which supplies only a small set of extracted memories. Presenting a large undifferentiated prompt dilutes relevant facts among irrelevant turns. Compression precision therefore determines accuracy more than context volume, and the full-context result represents the ceiling achievable without compression, a ceiling that selective extraction exceeds. \par

Only \gls{RAG} and mem0 are non-dominated on the statistical Pareto frontier. \gls{RAG} matches top-cluster accuracy at a \gls{TCO} of \$0.65, requiring no \gls{LLM} calls during loading and no graph infrastructure. Mem0 achieves the highest observed accuracy (81.08\%) at \$5.43, below both full-context (\$10.22) and Graphiti (\$6.95). Full-context is dominated simultaneously on cost and accuracy. Graphiti and cognee are dominated on accuracy at comparable or higher cost. For cost-sensitive \gls{IoA} deployments, \gls{RAG} is the pragmatic default, and mem0 is appropriate where accuracy gains justify the expenditure. \par

Accuracy and unknown rate are invariant to network constraints for every backend (all $p{>}0.07$, $\alpha{=}0.01$). The memory index content, established during loading, determines what the agent can answer, and the link over which memories are subsequently fetched does not. A statistically significant latency increase of 4\% to 5\% is observed for mem0 and \gls{RAG} under the constrained scenario (both $p = 0.006$), as both retrieve context across the edge-cloud link at query time. Full-context, Graphiti, and cognee show no significant latency response, their processing being compute-bound at the tested constraint level. Link quality governs what a \gls{DMAS} pays in latency, but not what it can remember. \par

These results diverge substantially from both vendor-published evaluations, illustrating the evaluation bias identified in Section~\ref{sec:introduction}. Mem0's published benchmark reports base accuracy of 66.88\% on LoCoMo~\cite{chhikara_mem0_2025}, versus 81.08\% found here. Zep's counter-evaluation reports Graphiti at 75.14\% and mem0 at 66.88\%~\cite{chalef_is_2025}, directionally opposite to the present findings. The testbed here uses evaluation code from both authors' repositories and the Graphiti-derived judge prompt, which should if anything favor Graphiti. Despite this, Graphiti's 25.58\% unknown rate persists across all ten conversations, pointing to indexing incompleteness as the cause rather than any evaluator artefact. The mutual disparity confirms that a shared harness, judge, infrastructure, and dataset are necessary for cross-framework comparison, and that neither vendor benchmark satisfies all four simultaneously. \par

\textbf{\hyperref[rq:1]{RQ1}} is answered by the testbed presented in Section \ref{sec:method}. Reproducible \gls{LTM} comparison in a \gls{DMAS} requires network isolation across separated subnets, deterministic network shaping, unified \gls{LLM} routing and observability for normalized cost attribution, full-system resource instrumentation, and a majority-voted \gls{LLM}-as-a-judge with a shared prompt and zero \gls{LLM} temperature across all frameworks. A coordinator further enforces the order of treatments, and structures their outcomes. \par

In response to \textbf{\hyperref[rq:2]{RQ2}}, \hyperref[h:02]{$H_0^{(2)}$} cannot be rejected. Accuracy and unknown rate are statistically invariant to network conditions for all frameworks and baselines. This indicates that memory quality is primarily determined at ingestion time rather than at retrieval time. The transition to a constrained cloud-edge \gls{DMAS} introduces only a modest, architecture-dependent latency increase, affecting systems that perform query-time retrieval but leaving correctness unchanged. In this setting, network conditions act as a throughput constraint rather than an information constraint. \par

Regarding \textbf{\hyperref[rq:3]{RQ3}}, \hyperref[h:03]{$H_0^{(3)}$} is rejected ($p = 0.002$ for all cross-cluster comparisons). \gls{RAG} and mem0 are the only non-dominated systems on the joint cost-accuracy frontier. The dominance structure is primarily driven by retrieval coverage: systems with higher effective recall at ingestion or indexing time consistently outperform those relying on richer structural representations that are not fully recovered at query time. \gls{RAG} offers the most cost-efficient configuration, while mem0 provides the highest observed accuracy, with both occupying distinct optimal points on the cost-accuracy trade-off curve. \par

\subsection*{Limitations}

This section outlines the four key limitations of the study, including constraints in benchmark coverage, system selection, testbed design, and ingestion granularity that bound the generalizability of the results. \par

A key limitation of this study is the reliance on the LoCoMo benchmark without evaluation across additional \gls{LTM} benchmarks, despite prior work identifying several methodological shortcomings. LongMemEval and ES-MemEval note that LoCoMo does not adequately evaluate reasoning over updated user states and lacks conflict detection, abstention behavior, and explicit user modeling \cite{wu_longmemeval_2025, chen_es-memeval_2026}. MobileMem highlights its single-session structure, limiting multi-session goal continuity, while MemoryAgentBench distinguishes long-context evaluation from true \gls{LTM}, where memory should be a compressed and selective representation rather than a full transcript \cite{deng_mobilemem_2026, hu_evaluating_2025}. Zep further critiques its limited difficulty due to context windows that allow strong full-context baselines. Additional issues include incomplete ground truth, multimodal inconsistencies, and ambiguous questions \cite{chalef_is_2025}. \par

A second limitation in the comparison of \gls{LTM} is the lack of intra-class evaluation, where only a single representative system is included for each category. In particular, the study does not include multiple vector-based, graph-based, and hybrid \gls{LTM} frameworks, limiting the ability to assess variance within each architectural class or to determine whether observed effects are method-specific or representative of the broader class behavior. \par

A third limitation concerns the design of the testbed. The system is deployed on a single machine inside Docker, where all components share the same underlying computational resources. The responder acts as a central point through which each Q\&A query from the coordinator is routed, making the system a quasi-distributed \gls{MAS} rather than a \gls{DMAS} by definition. Although bandwidth constraints, increased latency, and injected jitter were introduced to approximate a \gls{DMAS} environment, the resulting differences between constrained and unconstrained scenarios were not statistically significant. \par

A fourth limitation concerns the evaluation of knowledge ingestion strategies. The study focuses on ingestion at the granularity of individual conversational turns, which reflects episodic interaction data but does not cover alternative ingestion approaches such as long-form documents or large unstructured corpora. This is particularly relevant for cognee, which is designed to operate on extensive inputs including full documents and large datasets rather than isolated dialogue segments. \par

\subsection*{Future Work}

In the field of \gls{LTM}, future work could explore the deliberate fragmentation or partial removal of stored memory after distribution, either in vector space representations or within segments of knowledge graphs, to evaluate how effectively \gls{LTM} systems can recover, reconstruct, or infer missing information. This would provide a more direct measure of robustness under incomplete memory conditions and help characterize the resilience of retrieval and reasoning mechanisms when confronted with partial or corrupted memory states. \par

Regarding \gls{DMAS} and the \gls{IoA}, future work could involve constructing an \gls{IoA} that integrates agent discovery, identity management, distributed architectures, edge deployments, and \gls{LTM} across physically distributed hardware. Experimental conditions should reflect real-world network characteristics, including measured latency, bandwidth constraints, and jitter, rather than simulated scenarios. Following established practices in distributed systems research, memory could be partitioned across agents using strategies such as random assignment, batching, or round-robin distribution, enabling systematic analysis of memory consolidation mechanisms across distributed components, including how distributed \gls{LTM} is merged, reconciled, and maintained consistently across agents. \par

Future studies of \gls{LTM} in \gls{DMAS} and the \gls{IoA} could further incorporate dynamic and adversarial conditions, including agents joining and leaving the network, and Byzantine fault tolerance scenarios. This would enable a more rigorous evaluation of robustness, consistency, and resilience of memory in \gls{DMAS} and the \gls{IoA} under realistic and adversarial deployment settings. \par

\subsection*{Conclusion}

This paper presents an independent reproducible testbed for evaluating \gls{LTM} frameworks in a simulated cloud-edge \gls{DMAS}, benchmarking mem0, Graphiti, and cognee alongside \gls{RAG} and full-context baselines across ten LoCoMo conversations under unconstrained and constrained network scenarios. \par

Accuracy separates into two statistically distinct clusters ($p = 0.002$, paired Wilcoxon, Bonferroni-corrected). Mem0, \gls{RAG}, and full-context reach 77\% to 81\%, while Graphiti and cognee reach only 55\% to 56\%, a gap driven by retrieval incompleteness rather than reasoning failure. Compression precision determines accuracy more than context volume, as full-context forwarding underperforms mem0's selective extraction despite supplying the entire conversation per question. Network constraints leave retrieval quality unchanged, imposing a modest latency penalty on vector backends. \par

On the cost-accuracy frontier, only \gls{RAG} and mem0 are non-dominated. \gls{RAG} matches the upper cluster at 8.4 times lower \gls{TCO} than mem0, making it the pragmatic default for cost-sensitive \gls{IoA} deployments. Mem0 is preferable where maximum accuracy justifies the additional expenditure. \par

\section*{Acknowledgment}

The authors thank Prof. Emil Björnson for his mentorship during the writing of a university report that led to this publication, and Aleksey Veresov for his review of the manuscript. ChatGPT and GPT-5.2 were used to generate the code of the testbed. Claude Code and Opus 4.7 were used to extend the code of the testbed based on the reviewer’s comments. Cursor and Claude Sonnet 4.6 were used to generate the interactive Python notebook used to analyze the results. GPT-5.3-mini, GPT-5.5, Claude (Web) and Claude Opus 4.7 were used to incorporate the reviewer’s comments into the manuscript. \par

\bibliographystyle{IEEEtran}
\bibliography{Compsac-2026}

\end{document}